# Radiative Cooling and Thermoregulation in the Earth's Glow with Micropatterned Directional Emitters

Mathis Degeorges,[1,2] Jyothis Anand,[3] Anson Tsang,[1] Zhenpeng Li,[1] Nithin Jo Varghese,[1] Jyotirmoy Mandal[1,4]*

[1]Department of Civil & Environmental Engineering, Princeton University, Princeton, USA

[2]Institut National des Sciences Appliquées de Lyon, Lyon, France

[3]Buildings and Transportation Science Division, Oak Ridge National Laboratory, Oak Ridge, USA

[4]Princeton Materials Institute Princeton University, Princeton, USA

*Corresponding author: Jyotirmoy Mandal (jm3136@princeton.edu)

## Abstract

We demonstrate a micropatterned directional emitter (μDE) with an ultrabroadband, azimuthally selective and tailorable emittance across the thermal wavelengths and over wide angles. The μDE can enable a novel and passive seasonal thermoregulation of buildings by reducing summertime terrestrial radiative heat gain, and wintertime loss. We show several types of μDE, such as metallic, white and transparent variants, made using low-cost materials and scalable manufacturing techniques that are already in large-scale use. Furthermore, we show that its directional emittance can be geometrically tailored to sky-view factors in different urban scenarios. Outdoor experiments show that μDEs stay 1.53-3.26°C cooler than traditional omnidirectional building envelopes in warm weather, including when they are sunlit. In cold weather, μDEs can be up to 0.46°C warmer. Additionally, μDEs demonstrate significant cooling powers of up to 40 $Wm^{-2}$ in warm conditions and heating powers of up to 30 $Wm^{-2}$ in cool conditions, relative to typical building envelopes. Building energy models show that μDEs can achieve all-season energy savings similar to or higher than those of cool roofs. Collectively, our findings show μDEs as highly promising for thermoregulating buildings.

## Introduction

Thermoregulating living environments is an urgent challenge of our times, with implications across scales – achieving thermal comfort and energy savings in buildings,[1] reducing heat island effects,[2] and mitigating climate change by cooling localities and reducing $CO_2$ emissions.[3] To a large extent, thermal budgets of buildings and their environment are determined by radiative heat flows. Therefore, controlling them is key to addressing this challenge. Typically, radiative control is achieved with reflective envelopes that reduce heating in the solar wavelengths (λ~0.3-2.5 μm).[4,5] However, buildings also exchange much heat with the environment in the thermal infrared wavelengths (TIR, λ~2.5-40 μm). In this context, passive radiative cooling, which involves heat loss from terrestrial surfaces to space through the LWIR (λ~8-13 μm) atmospheric transmission window,[6,7] has recently gained prominence as a zero-energy, zero-carbon way to cool buildings, cities and larger environments.[1,2,8] Past works have demonstrated the use of photonic films,[9-11] paints,[5,12,13] and wood[14,15] for sub-ambient radiative cooling under sunlight. The zero-energy, zero-carbon functionality of these designs makes them highly attractive as a sustainable cooling option. Dynamic designs based on fluidic,[16] thermochromic,[17,18] and electrochromic[19] transitions, which are capable of passive seasonal thermoregulation, have also been reported. Collectively, these have marked major advances beyond traditional building envelopes like paint coatings,[20] glass and concrete, and have opened new possibilities for energy savings and thermal comfort.

However, a critical limitation of both traditional and emerging materials is their nearly omnidirectional thermal emittances ($\varepsilon$). This is because walls and windows, which often form most of a building's surface area, generally see a thermally oriented environment – warmer terrestrial features near and below the horizon, and the radiatively colder sky above (**Figure 1AFigure *1***). Typical emitters on vertical facades do lose heat to the sky, since their longwave radiance ($I_{emitter}$) exceeds downwelling atmospheric irradiance ($I_{sky}$).

However, their omnidirectional $\varepsilon$ also causes them to be heated by terrestrial irradiance ($I_{earth}$). Typically, the view-factor $v$ of the sky is $\leq 0.5$, and for perfect emitters at ambient temperature, the cooling potential takes the form:

$$P_{cooling} = \overbrace{v\,(I_{emitter} - I_{sky})}^{skywards\ heat\ loss} - \overbrace{(1 - v)\,(I_{earth} - I_{emitter})}^{terrestrial\ heat\ exchange} \quad (1)$$

While the skywards cooling is ~ 0-60 Wm$^{-2}$ for vertical facades depending on the weather, the ground can be hotter than 60°C in the summer (SI, Figure S16),[21,22] causing terrestrial heat gains of 50-100 Wm$^{-2}$. Therefore, $P_{cooling}$ is severely reduced, or even reversed, resulting in heating of vertical surfaces. In buildings, this increases temperatures, cooling loads, and greenhouse gas emissions, with the impact particularly intense in urban heat islands characterized by warm cityscapes and $v \ll 0.5$.

A dramatic reversal occurs during cold weather, when the colder ground acts as a heat sink (SI, Figure S16), causing buildings to radiate $\gtrsim$ 20 Wm$^{-2}$ heat to it and overcool. Thus, purely due to seasonal variations in $I_{earth}$, omnidirectional emitters on walls and windows cause overheating or cooling of buildings. Vertical facades often play a dominant role in the thermal budget of buildings, but this problem remains a major challenge.

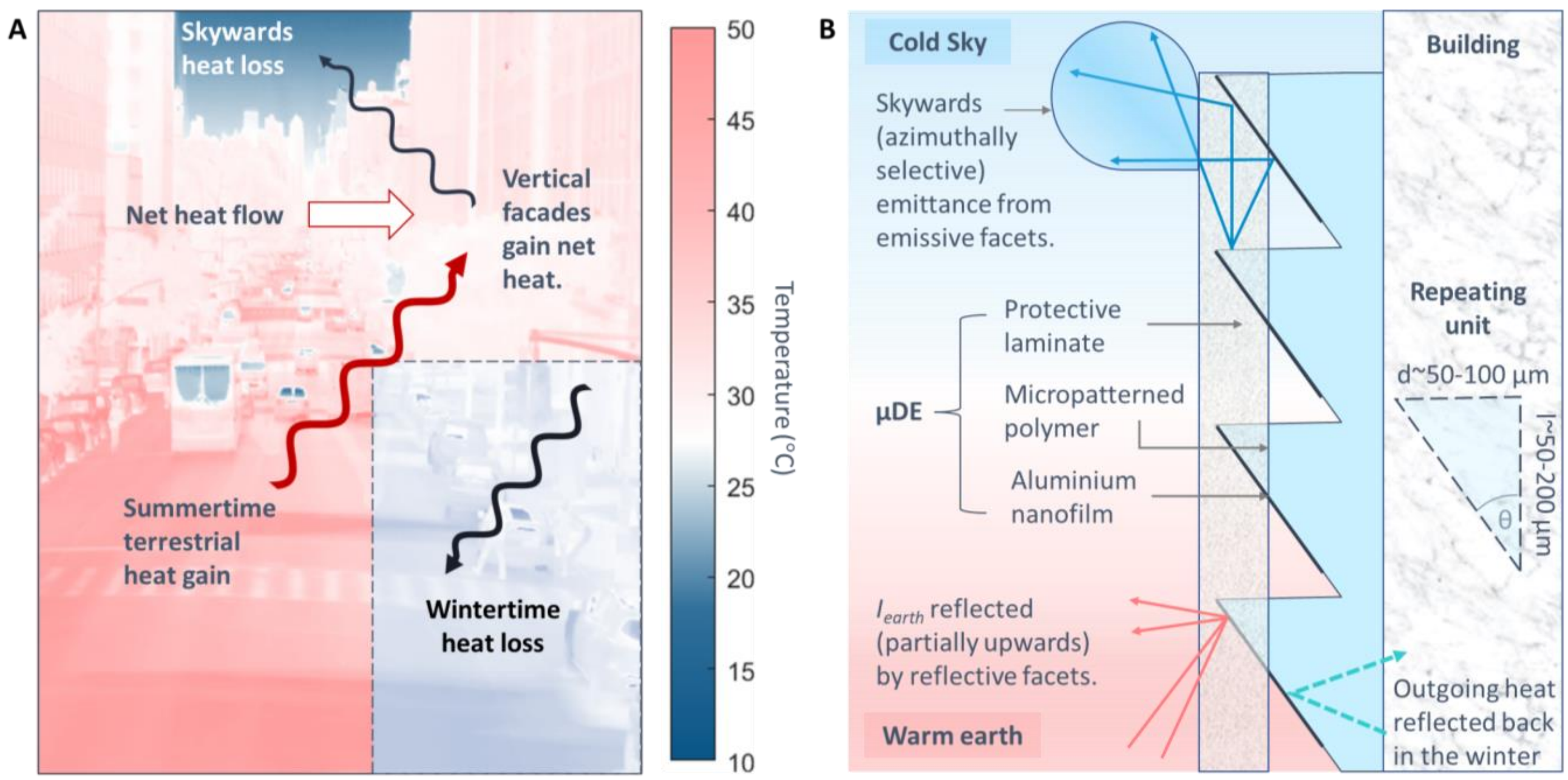

**Figure 1.** (**A**) Thermal image of an urban landscape with the radiative heat flows to which a facade is subjected, including undesired terrestrial heat gains in summer and heat losses in winter. (**B**) Schematic of our proposed solution, a sawtooth-micropatterned directional emitter (μDE), whose earth-wards facets are covered with metal to minimize radiative exchanges with the ground, and skywards facets are bare to maintain heat loss to the sky. The laminate reflects sunlight to keep the μDE cool, transmits TIR radiation, and offers protection from the elements.

Fundamentally, these limitations arise because traditional and emerging building envelope designs have not been tuned to thermally oriented environments. Indeed, recently reported radiative coolers and adaptive emitters have been mainly designed for sky-facing applications. In fact, vertically oriented designs often overlook the effects of terrestrial irradiance.[17,23] In a prior work, we had proposed the use of spectrally selective long-wave infrared (LWIR, λ~8-13 μm) emitters to address this issue.[24] However, the only established way to minimize terrestrial heat gains and losses is to reduce $\varepsilon$ altogether. Envelopes like metal sheets, low-$\varepsilon$ glasses,[25,26] and recently proposed colored variants[27] have high longwave reflectances that can reduce terrestrial heat gain. However, a low $\varepsilon$ also reduces skywards radiative cooling, and along with the considerable solar absorptance of these designs, traps additional solar heat.[28]

To address this longstanding issue, we report a micropatterned directional emitter (μDE) (**Figure 1B**) with azimuthally selective emittance ($\varepsilon$), which radiates heat upwards to space through the LWIR atmospheric window, and reflectively blocks radiative heat flows at angles near or below the horizon, reducing summertime terrestrial heat gain and wintertime loss. While two past works,[29,30] including one by us, have shown similar designs for cooling, to our knowledge, we are the first to report this novel and passive thermoregulation effect with directional emitters. In fact, past works have often regarded directional emitters as only suitable for cooling in warm weather, not for thermoregulation.[29] We demonstrate multiple embodiments of the concept, including silvery, white and transparent variants for walls and windows, with different geometries that yield different directional emittances, and using a range of common materials and highly scalable manufacturing techniques that are new in the context of directional emission (SI, Section 2, 7 and 8). Bare and porous polyethene (PE)-laminated embodiments of a specific μDE geometry exhibit above-horizon LWIR emittances $\varepsilon^{+}$of 0.93 and either 0.83 or 0.81, and below-horizon LWIR emittances $\varepsilon^{-}$ of 0.18 and 0.35 or 0.52, depending on the thickness of the porous polyethylene (**Figure 3F**, and SI, Section 2). The transparent variant we designed for windows also shows a highly directional $\varepsilon^{+}/\varepsilon^{-}$ combination of 0.79/0.44 (**Figure 4D,** and SI, Section 8). In outdoor tests, PE-laminated μDEs stay up to 1.53 or 3.26°C cooler than traditional omnidirectional emitters during summer days, depending on the thickness of the porous polyethylene, and 2.40°C cooler during summer nights. On winter nights, μDEs can be up to 0.46°C warmer (**Figure 5B**, and SI, Section 4, Table S2). We note here that the transparent embodiment that we show for windows is perhaps the first design that shows visible transmittance and TIR directionality. Preliminary building energy models show that μDEs can achieve all-season energy savings and $CO_2$ emissions reductions in buildings comparable to or greater than painting dark roofs white (SI, Section 6). Collectively, the findings show μDEs as a highly scalable design for achieving massive energy savings and thermal comfort in buildings.

## Micropatterned Directional Emitter (μDE) and Their Thermoregulation Capability

The μDE comprises a sawtooth patterned emitter with a microscale triangular repeating unit, and optionally, a solar reflective, TIR-transparent laminate (**Figure 1B**). The repeating units are made from a TIR-emissive dielectric, with the earth-facing facets coated with metal and the sky-facing facets bare. The intrinsic emissivity of the dielectric imparts a high $\varepsilon$ in the sky-facing direction, while the reflectivity of metal imparts a high reflectance towards the earth. The laminate, which is intrinsically non-absorptive, is nanostructured to effectively scatter and reflect sunlight, enhancing the μDE's cooling capability and giving it a similar appearance to traditional white facades, while acting as a TIR transparent effective medium that allows the directional $\varepsilon$ to be apparent.[16] It also serves as a protective layer.

The geometry of the μDE imparts an azimuthally selective $\varepsilon$ about the surface normal – high above the horizon, and low below, different from recent polar angle selective designs.[31] In the summer, this allows the μDE to lose heat skywards and reflect incident $I_{earth}$ to stay cooler than traditional omnidirectional emitters. In the winter, when the building is warmer than the cold ground, it reduces radiative heat loss relative to an omnidirectional emitter, keeping the building warm. This novel and seasonal thermoregulation is completely passive, and arises purely from the directional $\varepsilon$ and seasonality of terrestrial irradiance $I_{earth}$.

The μDE geometry arises from both fundamental and practical considerations. The choice of the sawtooth pattern is geometrically intuitive for a directional, skywards $\varepsilon$. The choice of ~50-200 μm features, too, is deliberate. Firstly, they are large enough to approach the geometric regime for ambient thermal radiation (λ~0.3-30 μm). This is crucial for simultaneous ultrabroadband and wide-angle control of $\varepsilon$ – which would be difficult for materials in the Mie scattering or metasurface regimes. At the same time, they are sufficiently small for the µDE to appear and feel nearly flat to human perception, which is important for use in facades.

The geometry ($l$, $d$, $\theta$) of the µDE can be altered to tailor the directionality and magnitude of $\varepsilon$ to optimize thermoregulation performance in both open ($v$~0.5, e.g. open spaces and suburbs) and congested ($v \ll 0.5$, e.g. scenarios with high-rise buildings) environments (SI, Figure S13). We have theoretically shown that the geometry can be tuned to continuously change between omnidirectional ($\varepsilon^+/\varepsilon^-$ = 1) and highly directional ($\varepsilon^+/\varepsilon^- \to \infty$) emittances (SI, Figures S22-23), and two experimental demonstrations of different geometries show reasonably good agreements with the theory (SI, Figure S24).

Given the variety of sky view-factors that vertical building facades may see, and the complexity of cooling demands, there is no ideal directional emitter that can achieve optimal cooling across all scenarios. Nonetheless, to show the thermoregulation potential of µDE, we consider the cooling performance of a step-directional emitter ($\varepsilon^+/\varepsilon^-$ =1/0), for different combinations of terrestrial and ambient air temperatures, dry (Total Precipitable Water, TPW = 10.5 mm) and extremely humid (TPW = 58.6 mm) conditions, and $v$=0.5. As evident from **Figure 2A**, when held at ambient temperature $T_{amb}$, a step DE can have cooling potential of as much as 120 Wm$^{-2}$ relative to an ideal omnidirectional emitter during summer days when ground temperature $T_{ground}$ is high, and 20 Wm$^{-2}$ heating potential during winter nights. **Figure 2B** shows analogous difference in steady state temperatures between a step DE and omnidirectional emitters, with cooling as high as ~7°C during summer days and warming of up to ~1°C during winter nights. Further details, including performance of the step DE and our fabricated µDEs, and the effect of view factor $v$, are presented in the Supporting Information, Section 3. Importantly, the relative cooling potential of µDEs are higher than in **Figure 2** when $v \ll 0.5$. Collectively, the results indicate that µDEs have a significant thermoregulation capability that could benefit buildings.

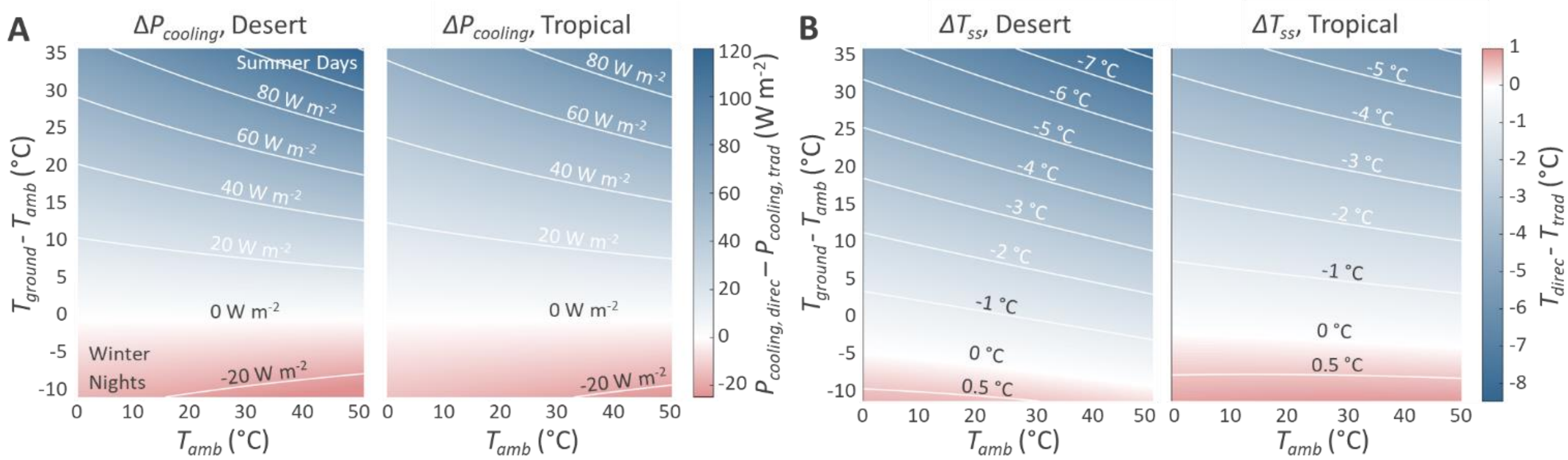

**Figure 2.** (**A**) $P_{cooling}$ differences between a step DE and an ideal omnidirectional emitter, as a function of $T_{ground}$ and $T_{amb}$ under desert (TPW = 10.5 mm) and extremely humid (TPW = 58.6 mm) conditions. Figure S9 of the SI shows cooling potentials of each emitter. (**B**) Steady state temperature differences between a step DE and an omnidirectional emitter. Figure S10 shows steady state temperatures of each emitter. An ideal solar reflectance ($R_{solar}$= 1) and a light wind ($h$ = 10 Wm$^{-2}$K$^{-1}$) is assumed. The transition from relative heating to cooling as the weather changes from cold to hot indicates the thermoregulation capability of directional emitters.

## Fabrication and Tests of Micropatterned Directional Emitter (µDE)

We designed our µDE as a film or cladding for walls and windows that is optically functional across the solar to far-IR wavelengths (λ~0.3-30 µm). The optical design requirements for the µDE can accommodate a wide range of materials and fabrication methods (SI, Section 1). The general requirement of a TIR-emissive dielectric is that most polymers, and ceramics like $SiO_2$ can serve as the emitter. The ~50-200 µm sawtooth pattern can be made using a number of highly scalable processes, such as thermal imprinting of thermoplastic polymers, photocuring of patterned ultraviolet-curable resins, and casting molten polymers or solvated ceramics on micropatterned molds. The solar reflective, TIR-transparent laminate can be made from porous, or zinc oxide, zinc sulfide or zinc selenide-doped polyethene.[32–34] The metal layer could be either aluminum or silver, and be deposited using physical vapor deposition techniques. These materials and methods are already made or used at very large scales for making both smooth and patterned films, often for use in or on buildings in other contexts.[35–41] We thus expect our μDEs to be highly scalable.

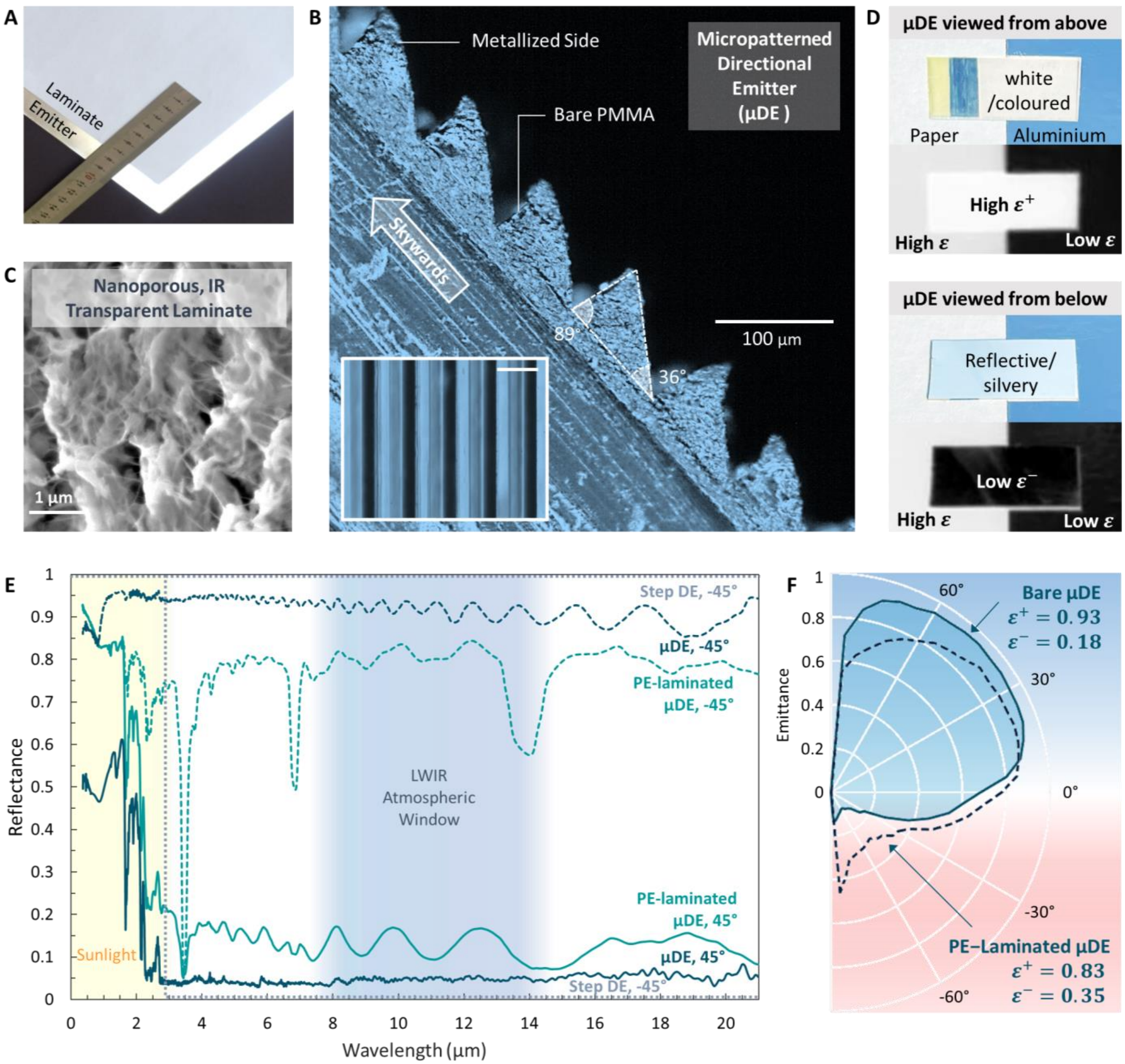

**Figure 3.** (**A**) Solar reflective μDE covered with porous polyethene laminate on top. (**B**) Microscopic image of the triangular micropattern where the metalized facets facing the ground, as well as the bare horizontal ones facing the sky, are identified. (**C**) SEM image of the polyethene laminate nanoporous structure. (**D**) Photos and LWIR thermographs of the μDE viewed from above and below, showing directional emittance. (**E**) Spectral reflectance of the step DE, and bare and PE-laminated μDE, at 45° above and below the horizon. (**F**) Emittance of the bare and PE-laminated μDE as a function of the angle relative to the horizon. Optical characterizations are detailed in the SI, Section 2.

To show the diverse possibilities, we fabricated several types of μDEs: acrylic and epoxy μDEs that are photocured against a micropatterned mold with UV light, polyethene, polypropene, and fluoropolymer μDEs thermally imprinted above their melting points. In what adds to the μDEs' environmental benefit, the polyethene and polypropene, were reused from waste. Section 1 and 8 of the Supporting Information shows the full set of examples. Here, we present a solar reflective variant for walls (**Figure 3A**), and a visibly transparent variant for windows (**Figure 4**). For the solar reflective variant, we chose a photocured acrylic μDE with aluminized (~100 nm thick) earth-facing facets, and a porous polyethene laminate on top. The photocuring involves a roll-to-roll fabrication (SI, Figure S1),[42] while the commercially procured porous polyethene laminate can be produced by phase inversion. Micrographs of the sawtooth micropattern and the laminate are shown in **Figures 3B** and **3C** respectively. As shown, the bare μDE has *l, d,* $\theta \sim$ 100 μm, 70 μm, 35° respectively. For the bare μDE, this leads to a highly directional $\varepsilon$, with integrated $\varepsilon^{+}$= 0.93 and $\varepsilon^{-}$= 0.18 (**Figure 3F**). The high $\varepsilon^{+}$ arises from the LWIR vibrational modes of the chemical bonds in the acrylic. The solar reflectance, measured at 45° above and below the horizon, is ~0.49 and 0.89 respectively (SI, Figure S6). The high value of reflectance below the horizon arises from light hitting the aluminized facets, while the lower value of reflectance above the horizon arises from light transmitted through the bare acrylic facets undergoing multiple reflections on the partially absorptive substrate beneath the μDE and the interior aluminum surfaces. Photo and LWIR thermographs confirm the μDE's directional $\varepsilon$ and solar transmittance (**Figure 3D**), with the former showing a way to achieve color as well.

With the porous PE laminate, the emittance contrast ($\varepsilon^{+}/\varepsilon^{-}$= 0.83/0.35 for a thin laminate, and 0.81/0.52, for a thick laminate, SI, Figure S4 and Table S1) is more subdued because of polyethene's small but omnidirectional emittance. However, the diffuse solar reflectance, arising from its nanoporosity (**Figure 3C**), is considerably higher (0.85 and 0.91 as opposed to 0.49, for the thin and thick laminates respectively), as required under strong sunlight (SI, Figure S6). The differences between the thick and thin laminates arise because a thicker porous polyethene, which reflects more sunlight by scattering, also absorbs more in the TIR wavelengths, where it acts as an effective medium.[16] We note that for PE laminates, higher solar reflectances and TIR transmittances may be possible,[38] and the color could also be altered using IR-transparent dyes and pigments to meet aesthetic requirements.[43] However, any colors must be light, as high solar absorption could make facades hotter than the terrestrial environment, in which case the μDE would be counterproductive.

The transparent μDE is a proof-of-concept design made from epoxy photocured against a mold (SI, Section 8). Based on our calculations *l, d,* $\theta \sim$ 290 μm, 80 μm, 15°, was chosen (SI, Figure S25). To maintain visible transmittance and high TIR reflectance, we fabricated a custom-made low- $\varepsilon$ coating, consisting of 40 nm of aluminum nitride, 14 nm of silver, and another 40 nm of aluminum nitride. A major challenge here was the refraction-induced turning of incident visible image due to the different textures of the back (flat) and front (sawtooth patterned) of the μDE (**Figure 4A** and SI, Figure S25B). We solved this using a transparent PE laminate (**Figure 4B**) that had almost the same refractive index as the epoxy (~1.50), and effectively made the structure into a planar and transmissive solid film in the visible (**Figure 4C** and SI, Figure S25C). However, the low-ε coating behaves as a reflector in the TIR, enabling a directional $\varepsilon$ around 30° above the horizon (**Figure 4D**), with $\varepsilon^{+}/\varepsilon^{-}$= 0.79/0.44. The visible specular transmittance, at near normal incidence, was 0.47 (**Figure 4E**). The spectral TIR reflectances are presented in **Figure 4F**.

To our knowledge, this is the first report of a planar, visible transmitter and directional TIR emitter concept. This is important because the directional emittance and its thermoregulation capability would have a large benefit on windows, which typically have low thermal resistances and are susceptible to parasitic heat gain.[45] We note that multilayer commercial low- $\varepsilon$ coatings[26] have significantly higher visible transmittance, solar reflectance, and TIR reflectance than the coating we designed (**Figure 4F**). Since it was not a focus of our paper, the low-$\varepsilon$ coating we created was used without optimization to demonstrate the proof-of-concept transparent μDE.

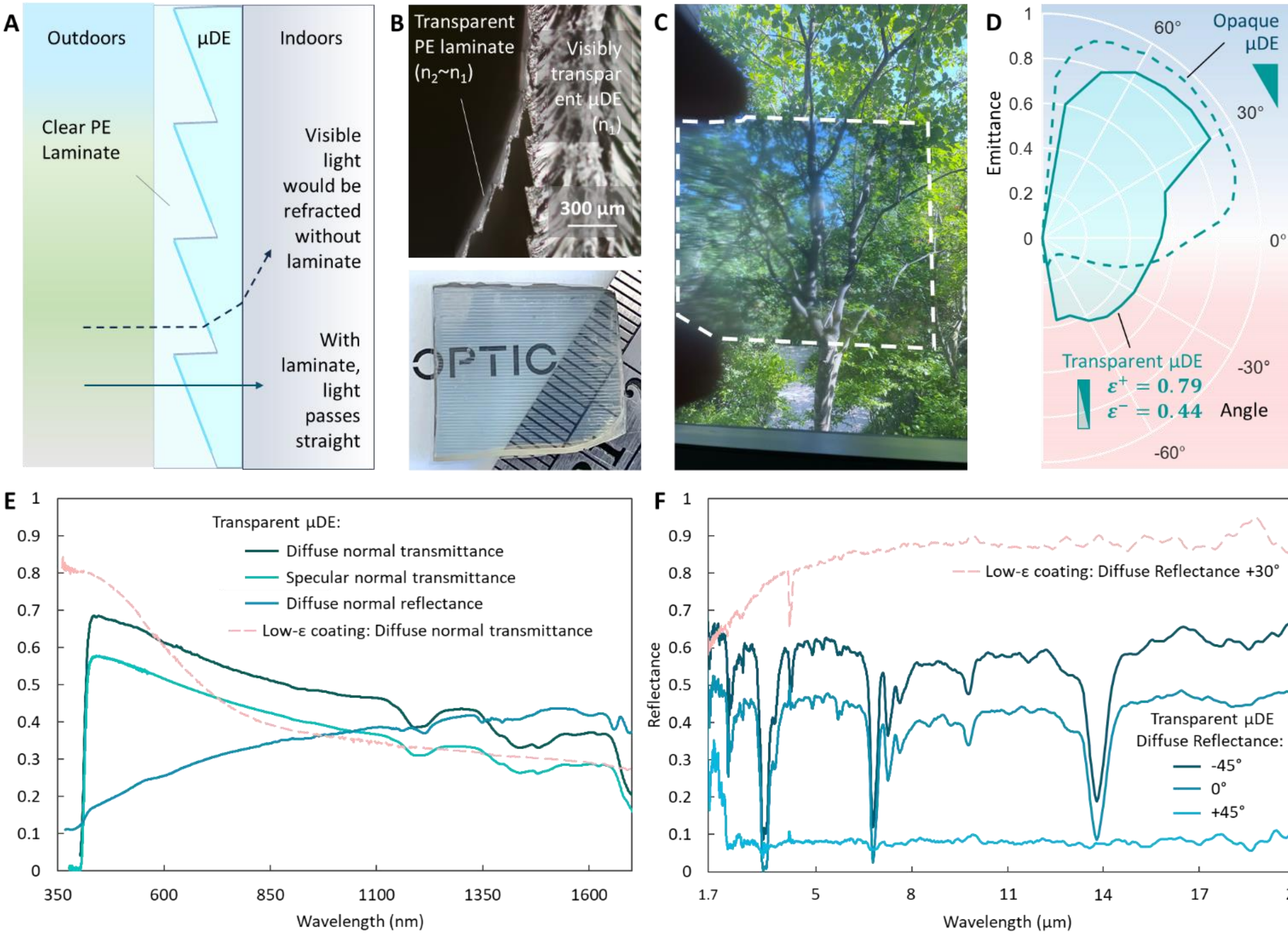

**Figure 4.** (**A**) Schematic of the transparent µDE, also featuring a sawtooth pattern, with a transparent PE laminate restoring the image that would be distorted by the combined effect of the pattern and the refractive index variation. (**B**) Micro and photograph of the transparent µDE sample with the PE laminate. (**C**) Image seen through the transparent µDE. (**D**) Emittance of the transparent µDE as a function of the angle relative to the horizon. The plot for the opaque µDE from Figure 3 is included for comparison. (**E**) UV-VIS-NIR specular transmittance, diffuse transmittance and reflectance of the transparent µDE and diffuse transmittance of the low-ε coating at normal incidence (**F**) TIR spectral diffuse reflectance of the transparent µDE, at normal incidence and at 45° below and above the horizon.

## Outdoor Performance Tests and Comparison Against Theory

To test the thermoregulation performance of the µDE, we measured its steady-state temperature relative to that of an omnidirectional control, a traditional white paint. Acrylic µDE and control samples, were mounted on a radiatively shielded R-26 insulation foam and exposed in a vertical orientation to ambient weather (**Figure 5A**) under clear skies, in both warm and cold weathers, and during the day and night. All experiments were done in Princeton, USA, either at street level or on roofs. The sky view factors ranged between ~0.4 and 0.5 (SI, Table S2).

**Figure 5B** shows a summary of results. Detailed results and analysis are presented in the SI, Section 4. During warm days characterized by hot ground and under the noontime sun, and under direct sunlight, porous PE-laminated acrylic µDEs with $R_{solar}$=0.85 and $\varepsilon^{+}/\varepsilon^{-}$= 0.83/0.35 stay 1.16-1.53 ± 0.10 cooler than a white painted omnidirectional emitter with $R_{solar}$= 0.86. The more reflective variant, with $R_{solar}$=0.91 and $\varepsilon^{+}/\varepsilon^{-}$= 0.81/0.52, stays 3.26 ± 0.10°C under a solar intensity of 673 Wm$^{-2}$. During warm nights, the PE-laminated µDE with $\varepsilon^{+}/\varepsilon^{-}$= 0.83/0.35 remains 2.40± 0.10°C cooler. On cold nights, when the ground is cooler than the ambient air, the bare and PE-laminated µDEs respectively stay 0.29 ± 0.10 and 0.46 ± 0.10°C warmer. Temperature-time plots for representative experiments are presented in **Figures 5C** and **D**. The ± 0.10°C uncertainty arises from the measured variation between thermocouples used in the experiments (SI, Section 4). Theoretical calculations performed using ambient solar, sky and terrestrial radiation measurements are largely consistent with the multiple experiments we performed, with small divergences attributable to uncertainties in measurements. We note that one experimental result (**Figure 5B**, Experiment 8) slightly outperforms our theoretical prediction, but present it for completeness and its qualitative consistency with the theory.

We also measured the μDE's cooling or heating power relative to an omnidirectional control in wintertime and simulated summertime conditions. The experimental configuration was the same, except for thermal loads attached to the back of the samples (SI, Section 4, Experiments 9 and 10). To eliminate the confounding effects of sunlight, the experiments were done at night. For the wintertime experiment, the μDE and control samples exposed were simultaneously heated by identical heaters to different powers and had their steady-state temperatures recorded (SI, Figure S14A). From the measurements, the differential heating power needed to maintain the samples at the same temperature relative to $T_{amb}$ was calculated. For the summertime experiments, μDE and control samples had a 1 $m^2$ heater placed in front of them to mimic the warm earth. The heater was covered with polyethene bubble wrap to prevent convective heating of the samples. The samples were connected to identical, flat containers of cold water otherwise thermally insulated from the environment (SI, Figure S15A). Temperatures of the water for both emitters were monitored as heat flowed through the emitters into them. The rate of heat gain for a given temperature relative to $T_{amb}$ was calculated for both the μDE and the control.

**Figures 5E** and **5F** show the differential heat flows for both experiments. Since the weather was very quiet for both experiments, the convective heat transfer coefficient was assumed constant throughout each experiment, meaning that the observed differences were radiative in origin. As shown, when held at 6°C and 10°C above $T_{amb}$, and 3°C and 7°C above the terrestrial environment (i.e. the environment below the horizon), the porous PE-laminated acrylic μDE loses 16 and 29 $Wm^{-2}$ less radiative heat than traditional white paint (**Figure 5F** and SI, Experiment 9). When 4°C and 7.75°C cooler than $T_{amb}$, and 26°C and 29.75°C cooler than the terrestrial environment, the μDE gains 25 and 40 $Wm^{-2}$ less radiative heat (**Figure 5E** and SI, Experiment 10). The differential heat losses and gains are substantial, and importantly, correspond to mild wintertime and summertime scenarios for buildings (SI, Section 4, Experiments 9 and 10). In reality, walls and windows of buildings may be much cooler in the summer or warmer in the winter than $T_{amb}$ and the terrestrial environment, and the differential heat flows would be even greater than the substantial values observed (SI, Section 5).

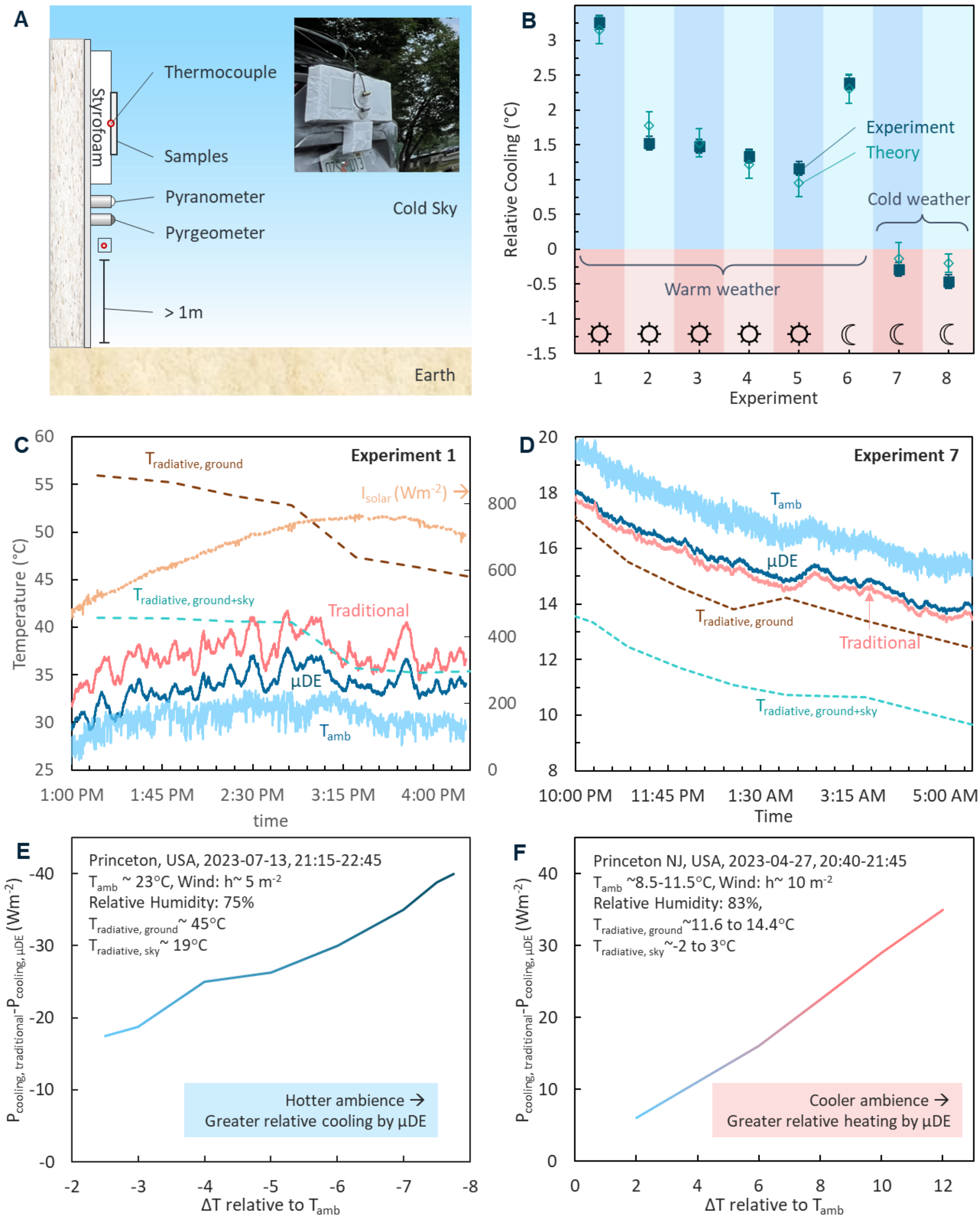

**Figure 5.** (**A**) Experimental setup for the steady state temperature measurement. Power flow measurements were similar, except for thermal loads attached to the back of the emitters. (**B**) Steady state temperature differences between the µDE and a traditional white paint in warm and cold weather. ☼ and ☾ indicate day and night. (**C**) Temperature time plots for experiment 1, showing ambient air temperature ($T_{amb}$), broadband ambient radiative temperatures of the environment ($T_{radiative,ground+sky}$) and below the horizon ($T_{radiative,ground}$), and temperatures of the traditional emitter ($T_{traditional}$) and PE-laminated µDE ($T_{directional}$) in a warm weather (**D**) Analogous figure for experiment 7, where the bare µDE ($T_{directional}$) is tested in a cold weather. (**E**) Heat gains prevented (i.e. Cooling) by the µDE in warm conditions,

and (**F**) Heat losses prevented (i.e. Heating) by the µDE in cold conditions. Both cooling and heating are relative to an omnidirectionally emissive paint film and are plotted against the emitters' temperature relative to $T_{amb}$. The experiments and results are detailed in the SI, Section 4.

## Potential Impact on Buildings

The passive seasonal thermoregulation capability of µDEs deployed on walls and windows could yield massive energy savings and thermal comfort in buildings. We altered a previously validated building energy model[46] to calculate the benefits of the µDE during peak summer and winter in desert and tropical conditions (SI, Section 6). Our model shows that by greatly reducing terrestrial heat flows to/from vertical facades, a step DE can lower peak summertime heat inflow by ~0.102 to 0.228 kWh m$^{-2}$ day$^{-1}$ and wintertime heat outflow by ~-0.006 to 0.02 kWh m$^{-2}$ day$^{-1}$ through R13 insulated walls and brick walls respectively. For windows, the reduction is ~0.091 to 0.416 kWh m$^{-2}$ day$^{-1}$. For the PE-laminated µDE ($\varepsilon^+/\varepsilon^-$= 0.83/0.35) the summertime heat inflow reduction is ~0.043 to 0.095 kWh m$^{-2}$ day$^{-1}$ and wintertime heat outflow reduction is ~0.024 to 0.060 kWh m$^{-2}$ day$^{-1}$ through R13 insulated walls and brick walls respectively.[24,45]

Promisingly, the values for the PE-laminated µDE are 1.4-3.2x of the *per m²* benefits of painting dark roofs white.[1] For windows, the benefits are 6.4-7.4x. Assuming that 15% of the vertical facades are windows, and the vertical surface to roof area ratio is 4, µDEs may lower cooling/heating loads in buildings by ~8-15x more than cool roofs, and in all seasons. Applying this ratio to previously reported energy savings and $CO_2$ reductions by cool roofs,[1,47] we estimate that on small to mid-sized buildings (roof area of 200 m$^2$ and facade area of 800 m$^2$), µDEs may cut $CO_2$ emissions by ~10-17 tons, and save ~US$ 3000-5500 annually depending on the insulation (SI, Section 6, Figure S21). Notably, these benefits would be additional to that of having a high solar reflectance, as the core functionality of the µDE is in the TIR.

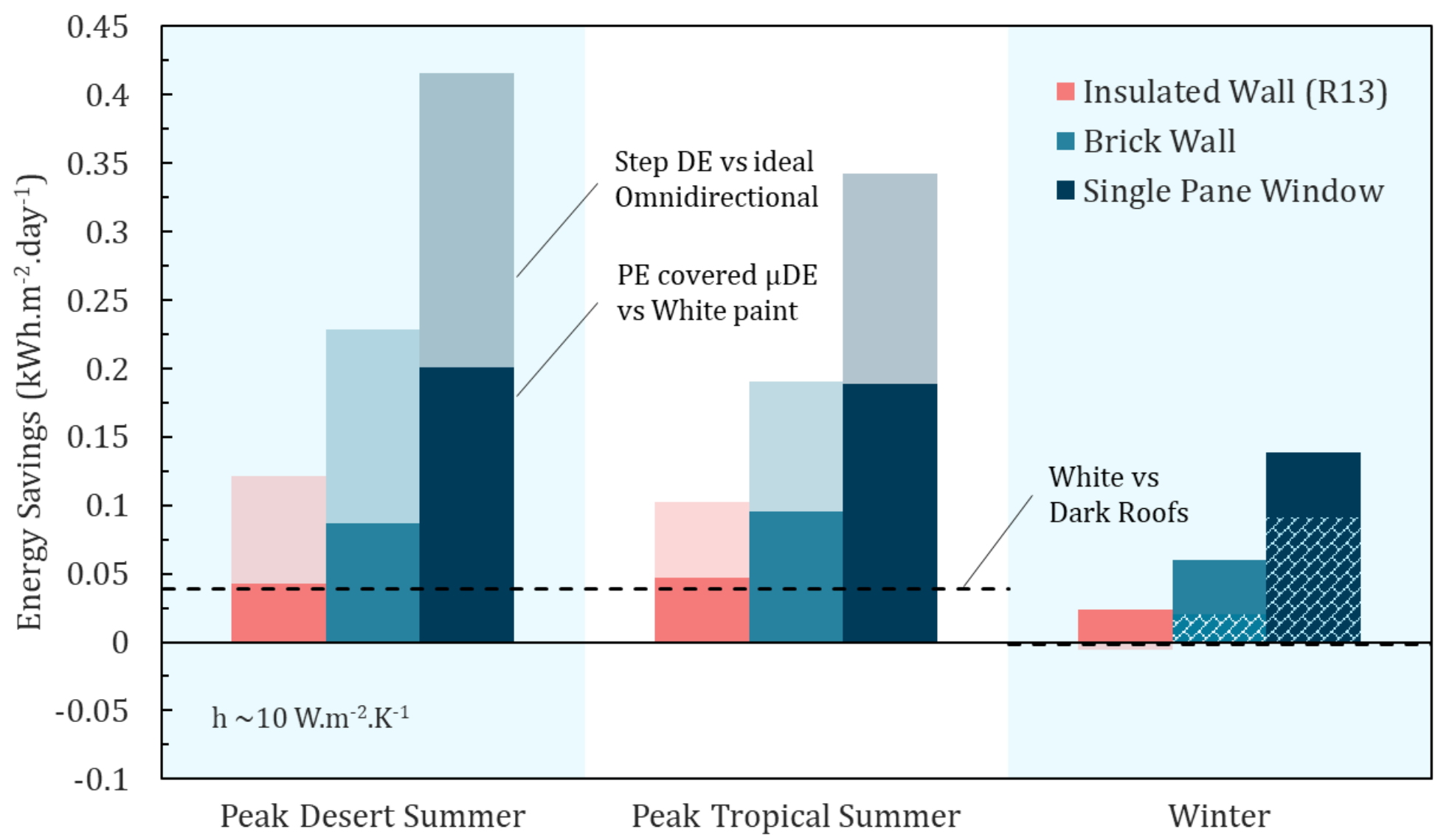

**Figure 6.** Energy savings achieved by using the µDE instead of a traditional omnidirectional emitter, in both peak desert and tropical summer and peak winter conditions, with a light wind (convective coefficient: $h$ = 10 W m$^{-2}$ K$^{-1}$). The three materials studied, the locations chosen, and all the details of the model used are presented in Section 6 of the Supporting Information. The doted lines correspond to the peak energy savings achieved by painting dark roofs in white.[1]

Lastly, we raise the possibility that µDEs may be able to cool urban canyons. While they behave similarly to white walls in the solar wavelengths by reflecting part of the $I_{earth}$ skywards, the combination of right µDE and urban canyon geometry[48] may effectively increase $v$ at the pedestrian level, preventing radiative heat trapping and lowering temperatures within the canyon. Besides increasing outdoor thermal comfort, this may indirectly lower air conditioning loads in buildings.[49]

## Conclusions and Outlook

In this work, we demonstrate a micropatterned directional emitter (μDE), whose emissive and reflective facets give it an ultrabroadband, azimuthally selective emittance across the thermal wavelengths and over wide angles. On buildings, the μDE can exhibit a novel and passive seasonal thermoregulation of buildings by reducing terrestrial radiative heat gain in the summer, and loss in the winter. Using common and low-cost materials and scalable processes that are already established as promising or suitable for buildings, we show several embodiments of the μDE, including metallic, white and transparent variants. Beyond materials and manufacturing techniques, we also show geometric tunability to tailor the directional emittance to sky-view factors expected in different urban scenarios. Outdoor experiments comparing μDEs with traditional building envelopes show theoretically consistent steady-state temperatures, and high cooling and heating powers. Preliminary building energy models show that μDEs can achieve building energy savings comparable to or exceeding those of cool roofs, while offering the benefits of both directional emittance and high solar reflectance.

Collectively, our findings show the μDE concept to be highly promising for application on buildings, and for future research. Potential explorations include the use of different materials and fabrication techniques for μDEs, and large-scale (e.g. small buildings) demonstrations of performance and weatherability in the field. Accurately modelling the energy savings and cooling impact of μDEs also requires urban climate and building energy models, which to accurately model their impact.

## Declaration of interests

The concept detailed in this paper was first publicly disclosed on arXiv on August 7, 2024. A provisional patent application (No.63/460,081) has been filed related to this article.